\documentclass[11pt]{article}
\usepackage{amsmath,amssymb,amsthm,amsxtra,overpic,bbm,bm,epsfig,ulem,color}
\def\thefootnote{\fnsymbol{footnote}}

\begin{document}

\vspace{0.2cm}

\begin{center}
{\Large\bf The effective neutrino mass of neutrinoless
double-beta decays: \\ how possible to fall into a well}
\end{center}

\vspace{0.2cm}

\begin{center}
{\bf Zhi-zhong Xing $^{a,c}$}
\quad {\bf Zhen-hua Zhao $^{b}$} \footnote{Corresponding author:
zhzhao@itp.ac.cn}
\\
{$^a$Institute of High Energy Physics $\&$ School of Physical Sciences,
University \\ of Chinese Academy of Sciences, Beijing 100049, China \\
$^b$Department of Physics, Liaoning Normal University, Dalian
116029, China \\
$^c$Center for High Energy Physics, Peking University, Beijing
100080, China}
\end{center}

\vspace{1.5cm}
\begin{abstract}
The neutrinoless double-beta ($0\nu 2\beta$) decay is currently the
only feasible process in particle and nuclear physics to probe whether
massive neutrinos are the Majorana fermions. If they are of the Majorana
nature and have a normal mass ordering, the effective neutrino mass term
$\langle m\rangle^{}_{ee}$ of a $0\nu 2\beta$ decay may suffer
significant cancellations among its three components and thus sink
into a decline, resulting in a ``well" in the three-dimensional
graph of $|\langle m\rangle^{}_{ee}|$ against the smallest neutrino
mass $m^{}_1$ and the relevant Majorana phase $\rho$. We present a
new and complete analytical understanding of the fine issues inside
such a well, and identify a novel threshold
of $|\langle m\rangle^{}_{ee}|$ in terms of the neutrino masses and flavor mixing
angles: $|\langle m\rangle^{}_{ee}|^{}_* = m^{}_3
\sin^2\theta^{}_{13}$ in connection with $\tan\theta^{}_{12} =
\sqrt{m^{}_1/m^{}_2}$ and $\rho =\pi$. This threshold point, which
links the {\it local} minimum and maximum of $|\langle
m\rangle^{}_{ee}|$, can be used to signify observability or
sensitivity of the future $0\nu 2\beta$-decay experiments. Given
current neutrino oscillation data, the possibility of $|\langle
m\rangle^{}_{ee}| < |\langle m\rangle^{}_{ee}|^{}_*$ is found to be
very small.
\end{abstract}

\begin{flushleft}
\hspace{0.8cm} PACS number(s): 14.60.Pq, 13.15.+g, 25.30.Pt
\end{flushleft}

\def\thefootnote{\arabic{footnote}}
\setcounter{footnote}{0}

\newpage


Since Ettore Majorana first formulated a fermionic particle that
should be its own antiparticle in 1937 \cite{Majorana},
a huge amount of attention has been paid to the
{\it Majorana fermions} in particle and nuclear physics and the
{\it Majorana zero modes} in solid-state physics \cite{R}. In
particular after the experimental discoveries of solar, atmospheric,
reactor and accelerator neutrino oscillations \cite{PDG},
whether massive neutrinos are the Majorana fermions becomes
an especially burning question among a number of fundamentally
important questions in neutrino physics and cosmology.
If this is the case, then the neutrinoless double-beta ($0\nu 2\beta$)
decays of some even-even nuclei are expected to take place \cite{Furry}.
Namely, $N(A, Z) \to N(A, Z+2) + 2 e^-$, where the lepton number
is violated by two units. Given the fact that the neutrino masses
are so small that all the lepton-number-violating processes must
be desperately suppressed, currently the unique and only feasible
way to demonstrate the Majorana nature of massive neutrinos is to observe
the $0\nu 2\beta$ decays. In this respect a number of ambitious
experiments are either underway or in preparation \cite{Giunti}.

In the standard scheme of three neutrino flavors the rate of a $0\nu 2\beta$
decay is proportional to the squared modulus of the effective
Majorana neutrino mass term \cite{2B}
\footnote{The phase convention taken here is highly advantageous when
considering the interesting and experimentally-allowed neutrino mass
limit $m^{}_1 \to 0$ (or $m^{}_3 \to 0$), in which $\rho$ (or
$\sigma$) automatically disappears \cite{XZ2015}.}
\begin{eqnarray}
\langle m\rangle^{}_{ee} = m^{}_1 |U^{}_{e 1}|^2 e^{{\rm i}\rho} +
m^{}_2 |U^{}_{e2}|^2 + m^{}_3 |U^{}_{e3}|^2 e^{{\rm i} \sigma} \; ,
\end{eqnarray}
where $m^{}_i$ denotes the $i$-th neutrino mass (for $i=1,2,3$),
$U^{}_{e i}$ is the corresponding element of the $3\times 3$
neutrino mixing matrix $U$ \cite{MNSP}, and $\rho$ and $\sigma$
stand for the Majorana phases. One often chooses to parametrize
$|U^{}_{e i}|$ as follows \cite{PDG}: $|U^{}_{e1}| = \cos
\theta^{}_{12} \cos\theta^{}_{13}$, $|U^{}_{e2}| = \sin
\theta^{}_{12} \cos\theta^{}_{13}$, and $|U^{}_{e3}| =
\sin\theta^{}_{13}$. The three mixing angles $\theta^{}_{12}$,
$\theta^{}_{13}$ and $\theta^{}_{23}$ have been determined to a good
degree of accuracy from current neutrino oscillation data, so have
been the value of $\Delta m^2_{21} \equiv m^2_2 - m^2_1$ and the
modulus of $\Delta m^2_{31} \equiv m^2_3 - m^2_1$ \cite{PDG}. But
the sign of $\Delta m^2_{31}$ and the two phase parameters in Eq.
(1) remain unknown, nor does the absolute neutrino mass scale. That
is why $|\langle m\rangle^{}_{ee}|$ is usually plotted as a function
of $m^{}_1$ in the normal mass ordering (NMO) case ($\Delta m^2_{31}
>0$) or $m^{}_3$ in the inverted mass ordering (IMO) case
($\Delta m^2_{31} <0$) by allowing $\rho$ and $\sigma$ to vary from
$0$ to $2\pi$ \cite{Vissani}. In such a so-called Vissani graph, a
two-dimensional ``well" can appear in the NMO situation due to a
significant cancellation among the three components of $\langle
m\rangle^{}_{ee}$. The bottom of the well signifies the case
of $|\langle m\rangle^{}_{ee}| \to 0$ \cite{Xing03}, a
disappointing possibility which is definitely consistent with the
present experimental data.

Two immediate questions are in order: (1) how possible for the three
neutrinos to have a NMO; (2) how possible for the actual value of
$|\langle m\rangle^{}_{ee}|$ to fall into the well and become
unobservable in any realistic $0\nu 2\beta$ experiments. A
combination of current atmospheric (Super-Kamiokande \cite{SK}) and
accelerator-based (T2K \cite{T2K} and NO$\nu$A \cite{NOVA}) neutrino
oscillation data preliminarily favors the NMO at the $2\sigma$
level. If this turns out to be the case, an answer to the second
question will be highly desirable because it can help interpret the
discovery or null result of a $0\nu 2\beta$ experiment in the
standard three-flavor scheme, although some kind of hypothetical
(ad hoc) new physics may also contribute to $|\langle m\rangle^{}_{ee}|$.

The present work aims to answer the second question by giving a new
and complete analytical understanding of the fine structure of the
three-dimensional well of $|\langle m\rangle^{}_{ee}|$ against
$m^{}_1$ and $\rho$, as illustrated in Fig. 1, where the best-fit
values $\Delta m^2_{21} = 7.54 \times 10^{-5}~{\rm eV}^2$, $\Delta
m^2_{31} = 2.47 \times 10^{-3}~{\rm eV}^2$, $\sin^2\theta^{}_{12} =
0.308$ and $\sin^2\theta^{}_{13} = 0.0234$ \cite{FIT} have been
taken as the typical inputs. We identify a novel threshold of
$|\langle m\rangle^{}_{ee}|$ which is located at the center of the
well: $|\langle m\rangle^{}_{ee}|^{}_* = m^{}_3
\sin^2\theta^{}_{13}$ in connection with $\tan\theta^{}_{12} =
\sqrt{m^{}_1/m^{}_2}$ and $\rho =\pi$. This threshold point links
the {\it local} minimum and maximum of $|\langle m\rangle^{}_{ee}|$,
and it can be used to signify the observability or sensitivity of
the future $0\nu 2\beta$-decay experiments. Given
current neutrino oscillation data, the possibility of $|\langle
m\rangle^{}_{ee}| < |\langle m\rangle^{}_{ee}|^{}_*$ is found to be
very small.
\begin{figure}
\centering
\includegraphics[width=5in]{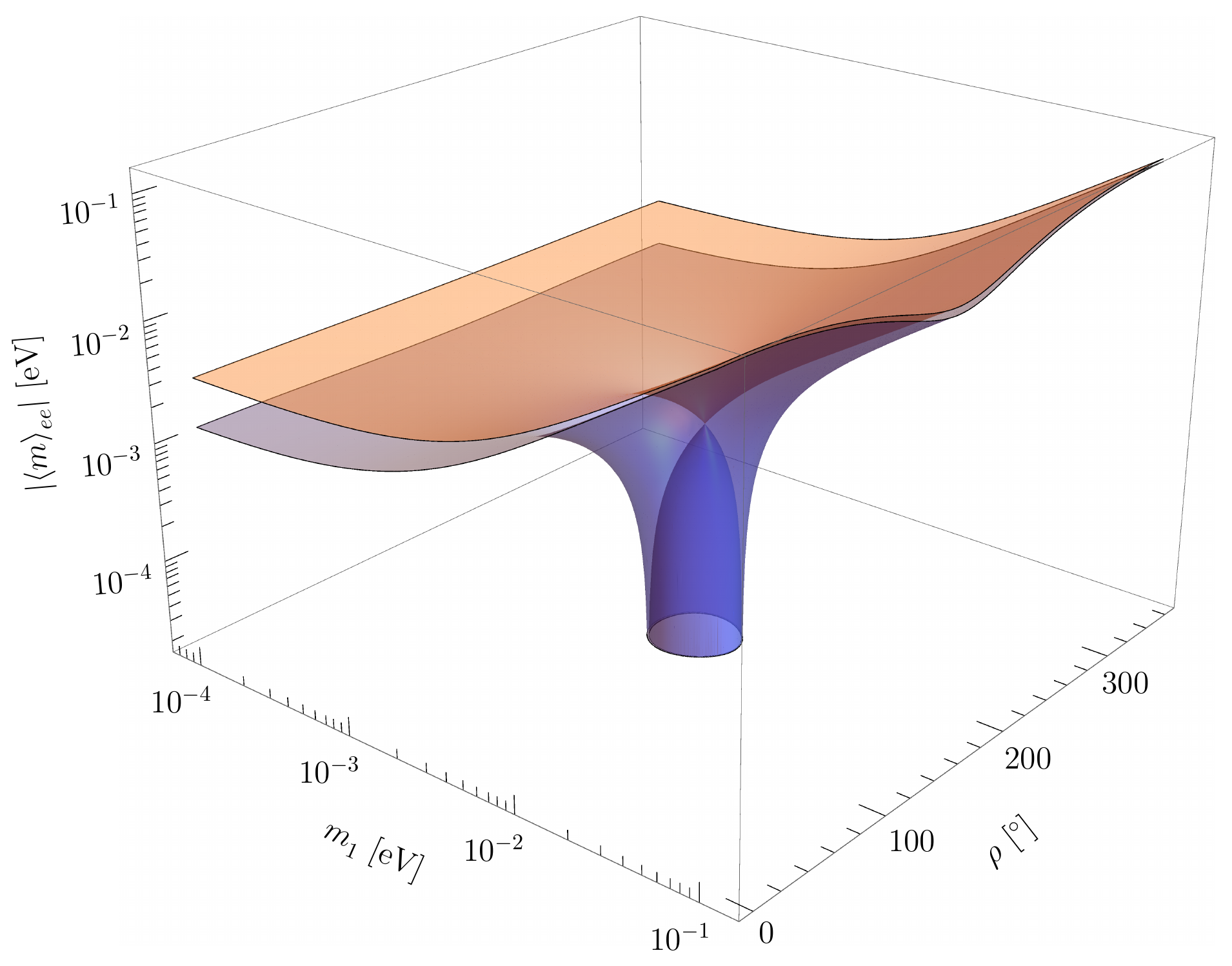}
\caption{Three-dimensional illustration of the upper (orange) and
lower (blue) bounds of $|\langle m \rangle^{}_{ee}|$ as functions of
$m^{}_1$ and $\rho$ in the NMO case, where the best-fit values
$\Delta m^2_{21} = 7.54 \times 10^{-5}~{\rm eV}^2$, $\Delta m^2_{31}
= 2.47 \times 10^{-3}~{\rm eV}^2$, $\sin^2\theta^{}_{12} = 0.308$
and $\sin^2\theta^{}_{13} = 0.0234$ \cite{FIT} have typically been
input.}
\end{figure}


Fig. 1 shows that the depth of the well of $|\langle
m\rangle^{}_{ee}|$ is mainly sensitive to a narrow parameter space
of $m^{}_1$ and $\rho$, while the other Majorana phase $\sigma$
plays an important role in shaping the bottom of the well
\cite{XZZ}. The latter point can be seen in an analytical way as
follows. Taking $\displaystyle\frac{\partial |\langle m
\rangle^{}_{ee}|}{\partial \sigma} =0$, we obtain
\begin{eqnarray}
\tan\sigma = \frac{m^{}_1 \sin\rho}{m^{}_1 \cos\rho +
m^{}_2 \tan^2\theta^{}_{12}} \; ,
\end{eqnarray}
so as to maximize or minimize $|\langle m \rangle^{}_{ee}|$ for the
given values of $m^{}_1$ and $\rho$. Substituting Eq. (2) into the
expression of $|\langle m \rangle^{}_{ee}|$ in Eq. (1), one arrives
at the following upper (``U") and lower (``L") bounds:
\begin{eqnarray}
\left|\langle m\rangle^{}_{ee}\right|^{}_{\rm U, L} = \left|
\overline{m}^{}_{12} \cos^2\theta^{}_{13} \pm m^{}_3
\sin^2\theta^{}_{13} \right| \; ,
\end{eqnarray}
where the sign ``$+$" (or ``$-$") corresponds to ``U" (or ``L"), and
\begin{eqnarray}
\overline{m}^{}_{12} \equiv \sqrt{m^2_1 \cos^4\theta^{}_{12} +
\frac{1}{2} m^{}_1 m^{}_2 \sin^2 2\theta^{}_{12} \cos\rho + m^2_2
\sin^4\theta^{}_{12}} \;\; .
\end{eqnarray}
It is easy to understand this result in an intuitive way: for any
given values of $m^{}_1$ and $\rho$, the maximum of $|\langle m
\rangle^{}_{ee}|$ comes out when the sum of the first two components
of $\langle m \rangle^{}_{ee}$ has the same phase as the third one
(i.e., $\sigma$); and the minimum of $|\langle m \rangle^{}_{ee}|$
arises when the difference between these two phases is equal to $\pm
\pi$. The bottom of the well shown in Fig. 1 corresponds to
$|\langle m\rangle^{}_{ee}|^{}_{\rm L} =0$, or equivalently
\begin{eqnarray}
\overline{m}^{}_{12} = m^{}_3 \tan^2\theta^{}_{13} \; .
\end{eqnarray}
Given the expressions $m^{}_2 = \sqrt{\displaystyle m^2_1 + \Delta
m^2_{21}}$ and $m^{}_3 = \sqrt{\displaystyle m^2_1 + \Delta
m^2_{31}}$ in the NMO case, Eq. (5) allows us to fix how the two
free parameters $m^{}_1$ and $\rho$ are correlated with each other.
Using the same best-fit inputs of $\Delta m^2_{21}$, $\Delta
m^2_{31}$, $\sin^2\theta^{}_{12}$ and $\sin^2\theta^{}_{13}$ as
those used in plotting Fig. 1, we illustrate the numerical
correlation between $m^{}_1$ and $\rho$ dictated by Eq. (5) in Fig.
2 --- the red curve. Such a correlation curve roughly looks like an
ellipse, but a careful analytical check shows that it does not
really obey the standard equation of an ellipse. Fig. 2 tells us
that touching the bottom of the well (i.e., $|\langle
m\rangle^{}_{ee}| \to 0$) is not a highly probable event at all,
because it requires $m^{}_1$ and $\rho$ to lie in the narrow regions
$2 ~{\rm meV} \lesssim m^{}_1 \lesssim 7 ~{\rm meV}$ and $0.86
\lesssim \rho/\pi \lesssim 1.14$, respectively \cite{Benato}.
\begin{figure}
\centering
\includegraphics[width=4in]{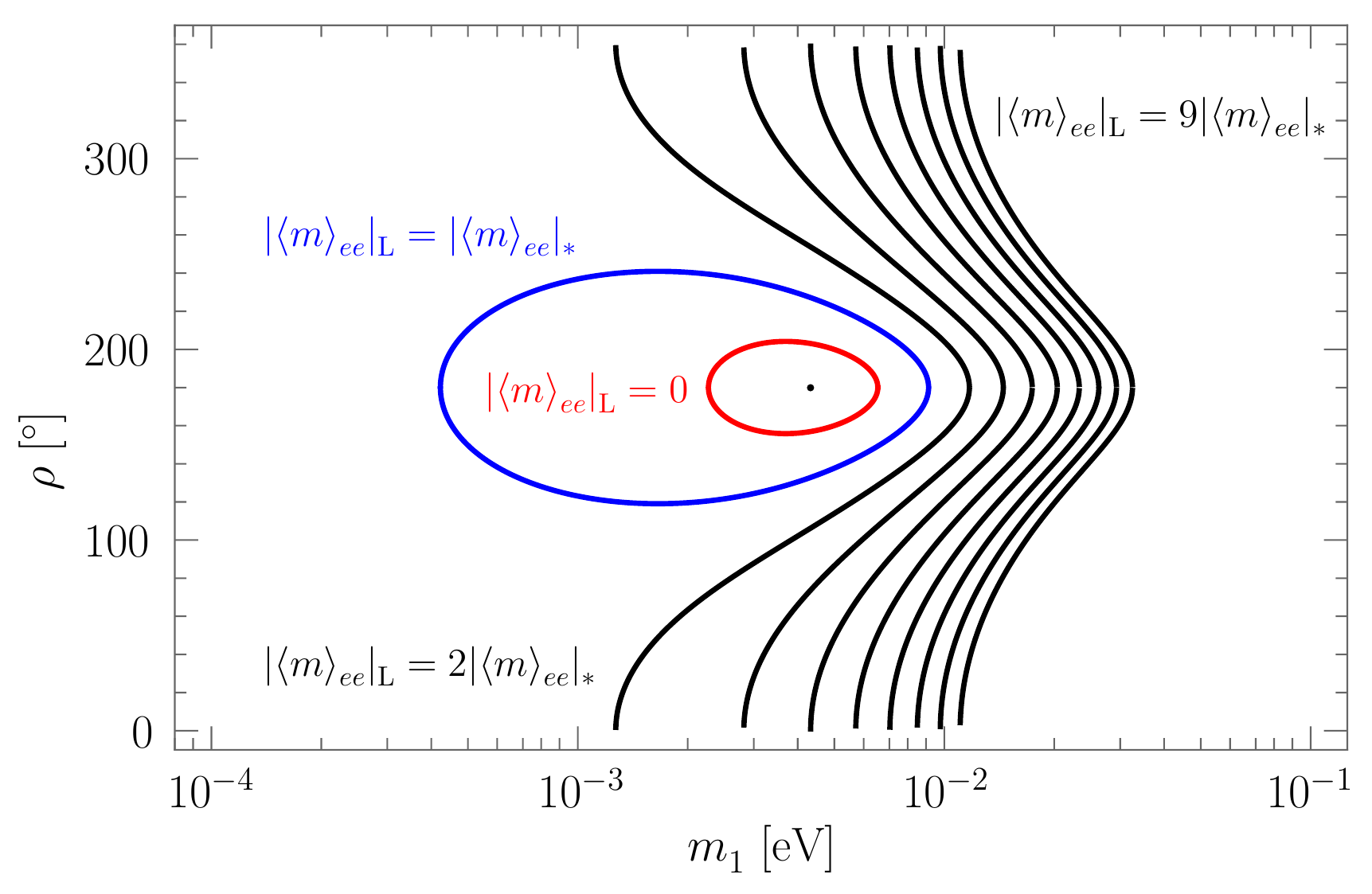}
\caption{The numerical correlation between $m^{}_1$ and $\rho$ in
three typical cases: (a) $|\langle m \rangle^{}_{ee}|^{}_{\rm L} =0$
(the red curve); (b) $|\langle m \rangle^{}_{ee}|^{}_{\rm L} =
|\langle m\rangle^{}_{ee}|^{}_{*} = m^{}_3 \sin^2\theta^{}_{13}$
(the black dot and the blue curve); and (c) $|\langle m
\rangle^{}_{ee}|^{}_{\rm L} = n |\langle m\rangle^{}_{ee}|^{}_{*}$
with $n \geq 2$ (the black curves). Here the best-fit values of
$\Delta m^2_{21}$, $\Delta m^2_{31}$, $\sin^2\theta^{}_{12}$ and
$\sin^2\theta^{}_{13}$ used in plotting Fig. 1 have been input.}
\end{figure}

Another salient feature of the well is the ``bullet"-like structure
of $|\langle m\rangle^{}_{ee}|^{}_{\rm L}$ as shown in Fig. 1,
corresponding to the parameter space of $\overline{m}^{}_{12} \leq
m^{}_3 \tan^2\theta^{}_{13}$. In other words, the surface of this
bullet is described by
\begin{eqnarray}
|\langle m\rangle^{}_{ee}|^{}_{\rm L} = m^{}_3 \sin^2\theta^{}_{13}
- \overline{m}^{}_{12} \cos^2\theta^{}_{13} \; .
\end{eqnarray}
The extremum of $|\langle m \rangle^{}_{ee}|^{}_{\rm L}$ in this inner
region of the well is supposed to be located at a point fixed by the
following two conditions:
\begin{eqnarray}
\frac{ \partial \left| \langle m \rangle^{}_{ee} \right|^{}_{\rm L} }
{\partial \rho} \hspace{-0.2cm} & = & \hspace{-0.2cm}
\frac{ m^{}_1 m^{}_2 \sin^2 2\theta^{}_{12} \cos^2\theta^{}_{13} }
{4 \overline{m}^{}_{12}} \sin\rho = 0 \; ,
\nonumber \\
\frac{ \partial \left| \langle m \rangle^{}_{ee} \right|^{}_{\rm L}
} {\partial m^{}_1} \hspace{-0.2cm} & = & \hspace{-0.2cm}
\frac{m^{}_1}{m^{}_3} \sin^2\theta^{}_{13} - \frac{m^{}_2
\overline{m}^2_{12} - \Delta m^{2}_{21} \sin^2\theta^{}_{12}
\left(m^{}_2 \sin^2\theta^{}_{12} + m^{}_1 \cos^2\theta^{}_{12}
\cos{\rho}\right)} {m^{}_1 m^{}_2 \overline{m}^{}_{12}}
\cos^2\theta^{}_{13} = 0 \; . \hspace{0.5cm}
\end{eqnarray}
The first condition definitely leads us to $\rho = 0$ or $\pi$. But
Fig. 2 clearly shows that $\rho$ should only take a value around
$\pi$ inside the well, and thus it is appropriate to take $\rho =
\pi$ instead of $\rho =0$. In this case $\overline{m}^{}_{12} =
|m^{}_1 \cos^2\theta^{}_{12} - m^{}_2 \sin^2\theta^{}_{12}|$ holds,
and the second condition in Eq. (7) is simplified to
\begin{eqnarray}
\frac{ \partial \left| \langle m \rangle^{}_{ee} \right|^{}_{\rm L}
} {\partial m^{}_1} = \frac{m^{}_1}{m^{}_3} \sin^2\theta^{}_{13} \pm
\left( \cos^2\theta^{}_{12} - \frac{m^{}_1} {m^{}_2}
\sin^2\theta^{}_{12} \right) \cos^2\theta^{}_{13} = 0 \; ,
\end{eqnarray}
where ``$\pm$" correspond to the prerequisites $m^{}_1 < m^{}_2
\tan^2\theta^{}_{12}$ and $m^{}_1 > m^{}_2 \tan^2\theta^{}_{12}$,
respectively. But in reality Eq. (8) can never be fulfilled since
its second term is much larger than its first term as a result of
(a) $2.50 \times 10^{-1} \leq \sin^2\theta^{}_{12} \leq 3.54 \times
10^{-1}$ and $1.85 \times 10^{-2} \leq \sin^2\theta^{}_{13} \leq
2.46 \times 10^{-2}$ at the $3\sigma$ level \cite{FIT} and (b)
$m^{}_1/m^{}_3 \leq m^{}_1/m^{}_2$ in the NMO case. Nevertheless,
Eq. (8) can at least allow us to draw a conclusion that is
absolutely consistent with current experimental data:
\begin{eqnarray}
\frac{ \partial \left| \langle m \rangle^{}_{ee} \right|^{}_{\rm L}
} { \partial m^{}_1 } \hspace{-0.2cm} & > & \hspace{-0.2cm} 0
\hspace{0.4cm} {\rm for} \hspace{0.4cm}  m^{}_1 < m^{}_2
\tan^2\theta^{}_{12} \; , \hspace{0.5cm}
\nonumber \\
\frac{ \partial \left| \langle m \rangle^{}_{ee} \right|^{}_{\rm L}
} { \partial m^{}_1 } \hspace{-0.2cm} & < & \hspace{-0.2cm} 0
\hspace{0.4cm} {\rm for} \hspace{0.4cm}  m^{}_1 > m^{}_2
\tan^2\theta^{}_{12} \; .
\end{eqnarray}
This observation means that $|\langle m \rangle^{}_{ee}|^{}_{\rm L}$
increases when $m^{}_1 < m^{}_2 \tan^2\theta^{}_{12}$ holds, and
it decreases when $m^{}_1 > m^{}_2 \tan^2\theta^{}_{12}$ holds.
Hence there must be a local maximum for $|\langle m
\rangle^{}_{ee}|^{}_{\rm L}$, denoted as
\begin{eqnarray}
|\langle m\rangle^{}_{ee}|^{}_{*} = m^{}_3 \sin^2\theta^{}_{13} =
\sqrt{m^2_1 + \Delta m^2_{31}} \ \sin^2\theta^{}_{13}
\end{eqnarray}
at the position fixed by $\rho = \pi$ and
\begin{eqnarray}
m^{}_1 = m^{}_2 \tan^2\theta^{}_{12} = \sqrt{m^2_1+\Delta m^2_{21}}
\tan^2\theta^{}_{12} \;\;\;\; {\Longrightarrow} \;\;\;\;
m^{}_1  = \sqrt{\Delta m^2_{21}} \ \frac{\sin^2\theta^{}_{12}}
{\sqrt{\displaystyle \cos 2\theta^{}_{12}}} \; .
\end{eqnarray}
In Fig. 1 this point is exactly the tip of the bullet inside the
well! In other words, the local maximum of $|\langle m
\rangle^{}_{ee}|^{}_{\rm L}$ arises from Eq. (6) at
$\overline{m}^{}_{12} =0$. Given the best-fit values of $\Delta
m^2_{21}$, $\Delta m^2_{31}$, $\sin^2\theta^{}_{12}$ and
$\sin^2\theta^{}_{13}$ that have been used in plotting Fig. 1, the
numerical location of the tip of the bullet turns out to be
$\left(m^{}_1, \rho, |\langle m\rangle^{}_{ee}|^{}_{*}\right) \simeq
\left(4 ~ {\rm meV}, 180^\circ, 1 ~{\rm meV}\right)$.

The above analysis explains why the bottom of the well does not
converge to a single point and why it is not flat either. In a
similar way one can understand why there is a local minimum for
$|\langle m\rangle^{}_{ee}|^{}_{\rm U}$, as shown in Fig. 1. The
extremum of $|\langle m\rangle^{}_{ee}|^{}_{\rm U}$ is expected to
be located at a position determined by
\begin{eqnarray}
\frac{ \partial \left| \langle m \rangle^{}_{ee} \right|^{}_{\rm U}
} {\partial \rho} \hspace{-0.2cm} & = & \hspace{-0.2cm} \frac{
m^{}_1 m^{}_2 \sin^2 2\theta^{}_{12} \cos^2\theta^{}_{13} } {4
\overline{m}^{}_{12}} \sin\rho = 0 \; ,
\nonumber \\
\frac{ \partial \left| \langle m \rangle^{}_{ee} \right|^{}_{\rm U}
} {\partial m^{}_1} \hspace{-0.2cm} & = & \hspace{-0.2cm}
\frac{m^{}_1}{m^{}_3} \sin^2\theta^{}_{13} + \frac{m^{}_2
\overline{m}^2_{12} - \Delta m^{2}_{21} \sin^2\theta^{}_{12}
\left(m^{}_2 \sin^2\theta^{}_{12} + m^{}_1 \cos^2\theta^{}_{12}
\cos{\rho}\right)} {m^{}_1 m^{}_2 \overline{m}^{}_{12}}
\cos^2\theta^{}_{13} = 0 \; . \hspace{0.8cm}
\end{eqnarray}
Of course, only $\rho = \pi$ is allowed with respect to the first
condition in Eq. (12). The second condition in Eq. (12) can never be
satisfied for the same realistic reasons given below Eq. (8). An
analogous and straightforward analysis tells us that the local
minimum of $|\langle m\rangle^{}_{ee}|^{}_{\rm U}$ exactly coincides
with the local maximum of $|\langle m\rangle^{}_{ee}|^{}_{\rm L}$,
and thus both of them are described by Eqs. (10) and (11). This
interesting result explains why the upper (in orange) and
lower (in blue) bounds of $|\langle m\rangle^{}_{ee}|$ connect with
each other in Fig. 1 when $m^{}_1 = \sqrt{m^2_1+ \Delta m^2_{21}}
\tan^2\theta^{}_{12}$ and
$\rho = \pi$ hold. Note that the overlap of the local maximum of
$|\langle m\rangle^{}_{ee}|^{}_{\rm L}$ and the local minimum of
$|\langle m\rangle^{}_{ee}|^{}_{\rm U}$ can also be understood from
Eq. (3) itself. At $m^{}_1 = \sqrt{m^2_1+ \Delta m^2_{21}}
\tan^2\theta^{}_{12}$ and $\rho
= \pi$, one simply has $|\langle m\rangle^{}_{ee}|^{}_{\rm L} =
|\langle m\rangle^{}_{ee}|^{}_{\rm U} = m^{}_3 \sin^2\theta^{}_{13}$
as a consequence of $\overline{m}^{}_{12} = 0$. So $|\langle
m\rangle^{}_{ee}|^{}_{*} = m^{}_3 \sin^2\theta^{}_{13} \simeq 1
~{\rm meV}$ stands for a threshold of $|\langle m\rangle^{}_{ee}|$
in the NMO case.

To visualize the steepness of the slope of $|\langle
m\rangle^{}_{ee}|^{}_{\rm L}$ around the well in Fig. 1, let us
project its contour onto the $m^{}_1$-$\rho$ plane by taking
$|\langle m\rangle^{}_{ee}|^{}_{\rm L} = n |\langle
m\rangle^{}_{ee}|^{}_{*}$ (for $n = 0, 1, 2, \cdots$) in Fig. 2. It
is especially interesting to compare between the contours of the
well at its bottom with $|\langle m\rangle^{}_{ee}|^{}_{\rm L} = 0$
(the red curve) and at its threshold height with $|\langle
m\rangle^{}_{ee}|^{}_{\rm L} = |\langle m\rangle^{}_{ee}|^{}_*$ (the
blue curve and the black point). They clearly show how the well
becomes narrower when the value of $|\langle
m\rangle^{}_{ee}|^{}_{\rm L}$ goes down. The profile of $|\langle
m\rangle^{}_{ee}|^{}_{\rm L}$ will be partially open and thus lose
its ``well" feature as $|\langle m\rangle^{}_{ee}|^{}_{\rm L} \geq 2
|\langle m\rangle^{}_{ee}|^{}_*$ is taken into account. Now that
$|\langle m\rangle^{}_{ee}|^{}_{\rm L} > |\langle
m\rangle^{}_{ee}|^{}_{*}$ always holds outside the blue curve in
Fig. 2, we argue that the parameter space of $|\langle
m\rangle^{}_{ee}|^{}_{\rm L} \leq |\langle m\rangle^{}_{ee}|^{}_{*}$
(i.e., $0.4 ~{\rm meV} \lesssim m^{}_1 \lesssim 10 ~{\rm meV}$ and
$0.66 \lesssim \rho/\pi \lesssim 1.34$) is a simple measure of the
chance for $|\langle m\rangle^{}_{ee}|$ to fall into the well and
become completely unobservable.

In general, $|\langle m\rangle^{}_{ee}|$ depends on all the three
unknown parameters $m^{}_1$, $\rho$ and $\sigma$. To illustrate how
probable or improbable for $|\langle m\rangle^{}_{ee}|$ to have a
value smaller than $|\langle m\rangle^{}_{ee}|^{}_{*}$ in a more
explicit way, we plot the three-dimensional parameter space of
$m^{}_1$, $\rho$ and $\sigma$ in Fig. 3, where the best-fit values
of $\Delta m^2_{21}$, $\Delta m^2_{31}$, $\sin^2\theta^{}_{12}$ and
$\sin^2\theta^{}_{13}$ used in plotting Figs. 1 and 2 have been
input. For clarity, the intersecting surfaces on the $\rho$-$\sigma$
plane corresponding to $m^{}_1 = 1, 2, 4$ and 6 meV are specified in
the figure. One can see that this parameter space is very small as
compared with the whole cubic space (i.e., the whole regions of
$m^{}_1$, $\rho$ and $\sigma$ allowed by current experimental
constraints). In comparison with $m^{}_1$ and $\rho$, the phase
$\sigma$ is only weakly constrained in Fig. 3. When the first two
components of $\langle m \rangle^{}_{ee}$ in Eq. (1) essentially
cancel each other out (i.e., $2 ~{\rm meV} \lesssim m^{}_1 \lesssim
7 ~{\rm meV}$ and $0.86 \lesssim \rho/\pi \lesssim 1.14$), a large
part of the range of $\sigma$ is allowed (e.g., the black
intersecting surface corresponding to $m^{}_1 = 4$ meV in Fig. 3).
But when the value of $m^{}_1$ decreases, the value of $\sigma$
should approach $\pi$, such as the green intersecting surface
corresponding to $m^{}_1 = 1$ meV in Fig. 3. In this case the second
component of $\langle m \rangle^{}_{ee}$ in Eq. (1) can be cancelled
by the other two components to a maximal level. For a similar
reason, the value of $\sigma$ should approach 0 or $2\pi$ when the
value of $m^{}_1$ increases (e.g., the blue intersecting surface
corresponding to $m^{}_1 = 6$ meV in Fig. 3). In any case we
conclude that the possibility of $|\langle m\rangle^{}_{ee}| <
|\langle m\rangle^{}_{ee}|^{}_*$ involves significant
cancellations among its three components and is really
small.
\begin{figure}
\centering
\includegraphics[width=4.8in]{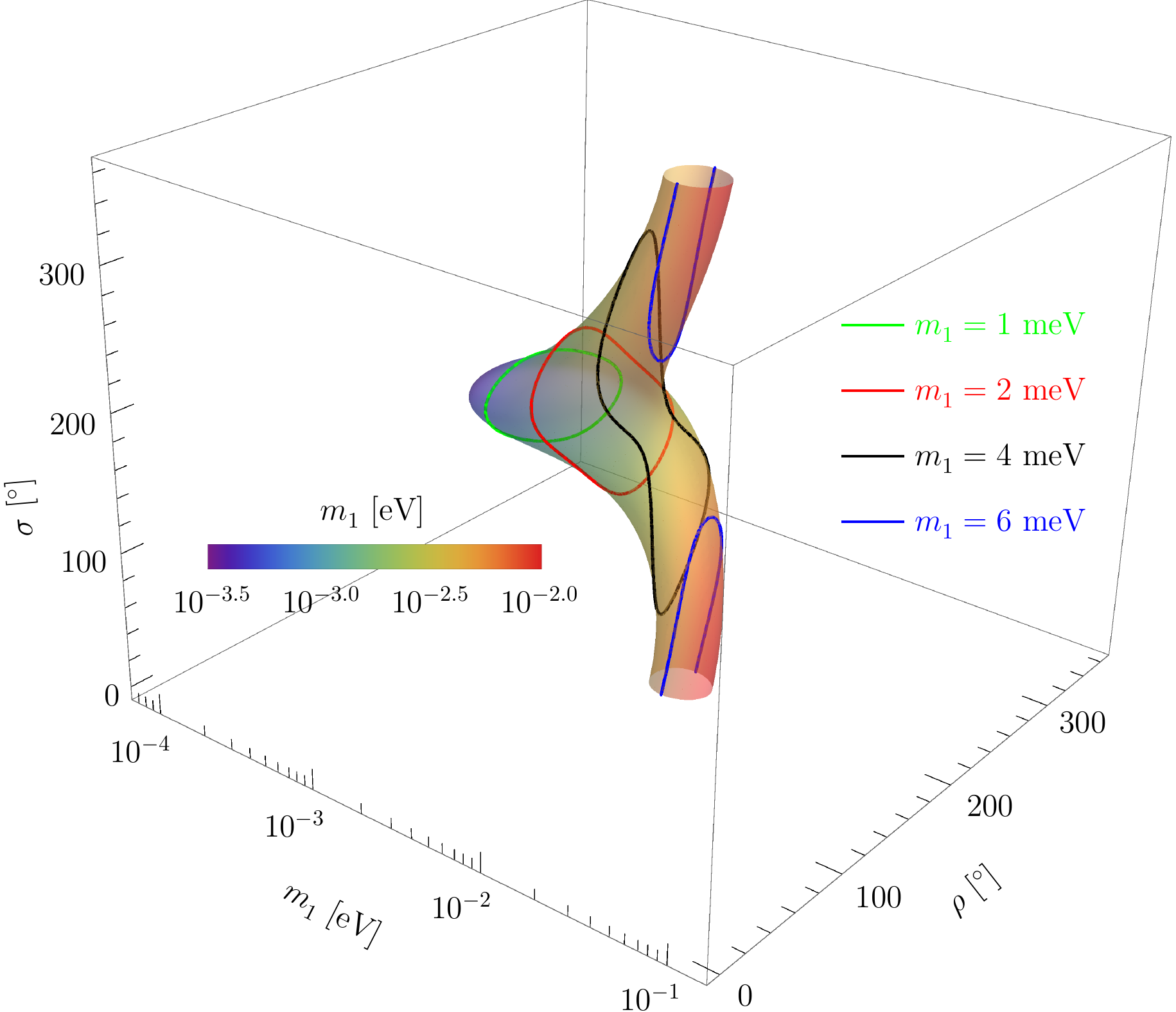}
\caption{The parameter space of $m^{}_1$, $\rho$ and $\sigma$
allowed for $|\langle m \rangle^{}_{ee}| < |\langle m
\rangle^{}_{ee}|^{}_*$ to hold, where the best-fit values of $\Delta
m^2_{21}$, $\Delta m^2_{31}$, $\sin^2\theta^{}_{12}$ and
$\sin^2\theta^{}_{13}$ used in plotting Fig. 1 have been input. The
intersecting surfaces for $m^{}_1 = 1, 2, 4$ and 6 meV on the
$\rho$-$\sigma$ plane are explicitly shown in the figure.}
\end{figure}

From an experimental point of view, the threshold
$|\langle m\rangle^{}_{ee}|^{}_{*}$ should signify
an ultimate limit of the reachable sensitivity to $|\langle
m\rangle^{}_{ee}|$ in the foreseeable future. At present the most
sensitive $0\nu 2\beta$-decay experiments can only set an upper limit of
$|\langle m\rangle^{}_{ee}|$ around $165 ~{\rm meV}$ \cite{K-Zen},
which depends on some theoretical uncertainties in calculating the
relevant nuclear matrix elements \cite{NME}. The most ambitious
next-generation high-sensitivity $0\nu 2\beta$-decay
experiments (e.g., nEXO \cite{NEXO})
are likely to probe $|\langle m\rangle^{}_{ee}|$ at the
level of a few tens of meV
\footnote{Note that the accuracy of a prediction for the experimental
sensitivity crucially depends on our knowledge of the relevant nuclear physics.
In the worst possible scenario, uncertainties from nuclear physics might even
weaken the expected experimental sensitivities by a factor as large as 5
\cite{Giunti}.}
\cite{Giunti}, a sensitivity still much larger than the threshold value
$|\langle m\rangle^{}_{ee}|^{}_{*} \simeq 1 ~{\rm meV}$
\footnote{In Ref. \cite{GL} a purely statistical analysis of the
possibility of $|\langle m\rangle^{}_{ee}| \lesssim 1$ meV has been
done to see to what extent the Majorana phases $\rho$ and $\sigma$
can be constrained for a given value of $m^{}_1$. While in Ref. \cite{PP}
the conditions for $|\langle m\rangle^{}_{ee}| > 1$ meV are
analyzed in the special case of $m^{}_1 \to 0$ or $\theta^{}_{13}
\to 0$. }.
In this sense there would be no hope to observe any $0\nu
2\beta$-decay signal if $|\langle m\rangle^{}_{ee}|$ were
unfortunately around or below the value of $|\langle
m\rangle^{}_{ee}|^{}_{*}$ in the standard three-flavor scheme.

Before ending our discussions about $\langle m\rangle^{}_{ee}$ and
its possible parameter space in the NMO case, let us briefly comment
on the relationship $\tan\theta^{}_{12} =\sqrt{m^{}_1/m^{}_2}$ from
a model-building point of view. This condition, together with $\rho
=\pi$, allows for $|\langle m\rangle^{}_{ee}| = |\langle
m\rangle^{}_{ee}|^{}_{*} = m^{}_3 \sin^2\theta^{}_{13}$ as a
remarkable threshold. It is well known that the Cabibbo angle
$\theta^{}_{\rm C}$ of quark flavor mixing can be related to the
ratio of quark masses $m^{}_d$ and $m^{}_s$ in a class of models
\cite{FX2000}: $\tan\theta^{}_{\rm C} \simeq \sqrt{m^{}_d/m^{}_s}~$,
which is consistent with the experimental data to a good degree of
accuracy. In comparison, the possibility of $\tan\theta^{}_{12}
\simeq \sqrt{m^{}_1/m^{}_2}$ is also interesting, in particular when
the NMO is true for the three mass eigenstates of $\nu^{}_e$,
$\nu^{}_\mu$ and $\nu^{}_\tau$ neutrinos. For example, we find that
an effective Majorana neutrino mass matrix of the form
\begin{eqnarray}
M^{}_\nu = \left( \begin{matrix}
0 & A & A \cr A & B & C \cr A & C & B \cr
\end{matrix} \right) -
m^{}_3 \frac{\sin\theta^{}_{13}}{\sqrt{2}} \left( \begin{matrix}
\sqrt{2} \ \sin\theta^{}_{13} & +{\rm i} & -{\rm i} \cr +{\rm i} & 0
& 0 \cr -{\rm i} & 0 & 0 \cr
\end{matrix} \right) \; ,
\end{eqnarray}
where $A$, $B$ and $C$ are all real, can essentially predict
$|\langle m\rangle^{}_{ee}| = m^{}_3 \sin^2\theta^{}_{13}$ and
$\tan\theta^{}_{12} = \sqrt{m^{}_1/m^{}_2}$ together with
$\theta^{}_{23} =\pi/4$, $\delta = -\pi/2$, $\rho = \pi$ and $\sigma
=0$ in the standard parametrization of $U$. Because $M^{}_\nu$
possesses the exact $\mu$-$\tau$ reflection symmetry, which can
easily be simplified to the $\mu$-$\tau$ permutation symmetry in the
$\theta^{}_{13} \to 0$ limit, one may take it as a starting point to
build a phenomenological neutrino mass model in this connection
\cite{XZ2016}.


In summary, we have achieved some new and important insights into the
effective neutrino mass $\langle m\rangle^{}_{ee}$ of the $0\nu 2\beta$
decays in the NMO case --- a case which seems to be more likely
than the IMO case according to today's preliminary experimental data.
Because $|\langle m\rangle^{}_{ee}|$ depends not only on
the unknown neutrino mass $m^{}_1$ but also on the free Majorana phases
$\rho$ and $\sigma$, a novel three-dimensional presentation of
$|\langle m\rangle^{}_{ee}|$ against $m^{}_1$ and $\rho$ reveals
an intriguing ``well" structure in the NMO case. The present work
provides a new and complete analytical understanding of the fine
issues inside such a well. We find a particularly interesting
threshold of $|\langle m\rangle^{}_{ee}|$ in terms of the neutrino
masses and flavor mixing angles:
$|\langle m\rangle^{}_{ee}|^{}_* = m^{}_3 \sin^2\theta^{}_{13}$
in connection with $\tan\theta^{}_{12} = \sqrt{m^{}_1/m^{}_2}$
and $\rho =\pi$. We suggest that this threshold point, which
links the local minimum and maximum of $|\langle m\rangle^{}_{ee}|$,
be used to signify observability or sensitivity of the future
$0\nu 2\beta$-decay experiments. In view of current neutrino
oscillation data, we conclude that the possibility of $|\langle
m\rangle^{}_{ee}| < |\langle m\rangle^{}_{ee}|^{}_*$ must be very
small. In other words, it should be very promising to detect
a signal of the $0\nu 2\beta$ decays and verify the Majorana nature
of massive neutrinos in a foreseeable future, even if they have
a normal mass spectrum.

One of us (Z.Z.X.) would like to thank J. Angel and S.T. Petcov for
interesting communications during the DBD16 workshop in Osaka, where
this work was initiated. We are also grateful to Y.F. Li, J. Zhang
and S. Zhou for some useful discussions, and to Z.C. Liu and Y. Lu
for their kind helps in plotting the figures. This work is supported
in part by the National Natural Science Foundation of China under
grant No. 11135009 (Z.Z.X.) and grant No. 11605081 (Z.H.Z.).

\end{document}